\documentclass{article}
\usepackage{graphicx}
\usepackage{amsmath} 
\usepackage{microtype} 
\usepackage[top=2.4cm, bottom=2.4cm, left=3cm, right=3cm]{geometry}

\let\OLDthebibliography\thebibliography
\renewcommand\thebibliography[1]{
  \OLDthebibliography{#1}
  \setlength{\parskip}{0pt}
  \setlength{\itemsep}{0pt plus 0.3ex}
}

\title{TUJU21: nuclear PDFs with electroweak-boson data at NNLO}

\author{Ilkka Helenius$^{1,2}$, Marina Walt$^{3,4}$ and Werner Vogelsang$^{3}$}

\date{%
  $^1$University of Jyvaskyla, Department of Physics, P.O. Box 35, FI-40014 University of Jyvaskyla, Finland \\
  $^2$Helsinki Institute of Physics, P.O. Box 64, FI-00014 University of Helsinki, Finland \\
  $^3$Institute for Theoretical Physics, University of Tübingen,
Auf der Morgenstelle 14, 72076 Tübingen, Germany \\
  $^4$HQS Quantum Simulations GmbH, Haid-
und-Neu-Straße 7, 76131 Karlsruhe, Germany \\
\vspace{6pt}
\emph{Presented at DIS2022: XXIX International Workshop on Deep-Inelastic Scattering and Related Subjects, Santiago de Compostela, Spain, May 2-6 2022}
}

\begin{document}

\maketitle

\section*{Abstract}

Nuclear parton distribution functions (nPDFs) can be determined in a global QCD analysis using a wide range of experimental data. In addition to older fixed-target deep inelastic scattering and Drell-Yan (DY) dilepton production data, several analyses from p+Pb collisions at the LHC provide further constraints and extend the kinematic reach of applicable data. Here we present an update of our previous TUJU19 analysis where we now include also electroweak-boson production data recently measured by ATLAS and CMS. For the first time, LHC data are included in a nPDF analysis performed at next-to-next-to-leading order (NNLO) in perturbative QCD. As before, our setup is based on the open-source analysis framework xFitter and we fit our own proton baseline, ensuring a fully consistent setup. We find good agreement with the applied data and that the resulting $\chi^2/N_{\mathrm{df}}$ is significantly smaller in case of the NNLO analysis (0.84) compared to our NLO analysis (0.94). Also, we present comparisons between our NNLO calculations and electroweak-boson production data in Pb+Pb collisions from ATLAS and CMS and DY data recently measured by CMS where NNLO corrections are found significant.

\section{Introduction}

Nuclear parton distribution functions (nPDFs) provide a framework to extract ``cold nuclear matter'' effects from measured data using a global DGLAP analysis and provide a baseline for collinear-factorization based cross-section calculations in high-energy heavy-ion collisions. Thanks to the many recent analyses performed by the different LHC experiments, the uncertainties of the nPDFs have been gradually shrinking and several assumptions needed earlier related e.g. to flavour dependence, have been relaxed. Within the last year all active groups have releases updated nPDF sets including varying amount of LHC data \cite{Duwentaster:2021ioo, Eskola:2021nhw, AbdulKhalek:2022fyi} where next-to-leading order (NLO) corrections in perturbative QCD (pQCD) are included in the cross section calculations and evolution equations. In case of proton PDFs, however, the precision of the data is already at the level where next-to-next-to-leading order (NNLO) computations are needed to match the precision of the applied observables. There have been a couple of nPDF analyses performed at NNLO accuracy but none of them has included any LHC data. Here we present the TUJU21 nPDF analysis \cite{Helenius:2021tof} where we update our TUJU19 nPDF analysis \cite{Walt:2019slu} by including recent ATLAS and CMS data for electro-weak boson production in p+Pb collisions at the LHC for the first time at NNLO.

\section{Analysis framework}

The fitting framework for the TUJU21 is based on the open-source tool \textsc{xFitter} \cite{xFitterDevelopersTeam:2017xal} which was extended to handle also nuclear PDFs in the context of our earlier TUJU19 nPDF analysis. One advantage of this setup is that we can easily fit our own proton baseline ensuring full consistency in the theoretical setup including parametrization scale, kinematical cuts and treatment of heavy-quark mass effects. We parametrize the proton PDFs using a similar form as in the HERA2.0 proton PDF analysis \cite{H1:2015ubc}
\begin{equation}
xf_{i}^{\,p}(x,Q^2_0) = c_0 x^{c_1} (1-x)^{c_2}(1 + c_3 x + c_4 x^2).
\label{eq:proton}
\end{equation}
To include nuclear effects, we follow a similar procedure as in nCTEQ15 and introduce an $A$ dependence for the proton PDF parameter as \cite{Kovarik:2015cma}
\begin{equation}
c_{k}(A) = c_{k,0} + c_{k,1}(1 - A^{-c_{k,2}}),
\label{eq:nuclear}
\end{equation}
where now parameters $c_{k,0}$ are obtained from the proton baseline analysis guaranteeing that in the $A \rightarrow 1$ limit the free proton PDFs are recovered. We parametrize separately the $\mathrm{u}$ and $\mathrm{d}$ quarks but in order to have convergent fits are required to assume no flavour dependence for the sea quarks, i.e. we set $\bar{\mathrm{u}} = \bar{\mathrm{d}} = \bar{\mathrm{s}} = \mathrm{s}$. We apply Hessian error analysis to quantify the resulting uncertainties with tolerance criteria $\Delta \chi^2 = 20~(50)$ for our proton (nuclear) fit and use FONNL-A(C) general-mass variable-flavour-number-scheme in the NLO (NNLO) fit to account for heavy-quark mass effects.

\section{Resulting PDFs}

\subsection{Proton baseline}

\begin{figure}[htb]
\begin{center}
\includegraphics[width=\textwidth]{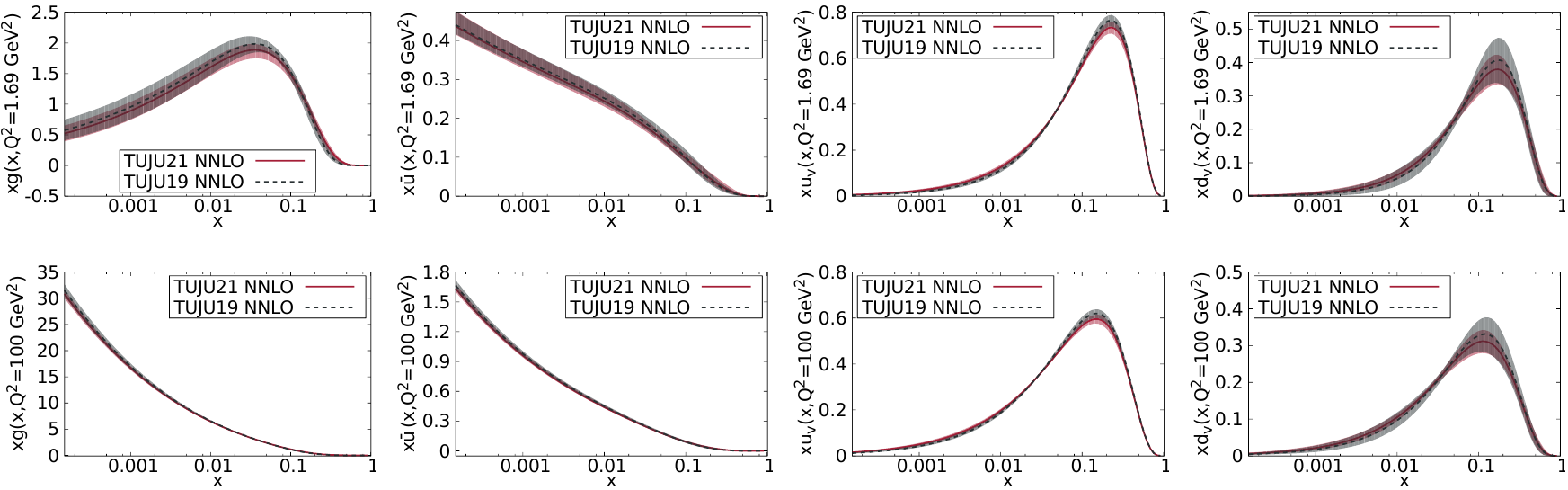}
\caption{Resulting proton baseline PDFs from TUJU21 analysis (solid red) and from our earlier TUJU19 analysis (dashed black) at NNLO with two different scales $Q^2 = 1.69~\textrm{GeV}^2$ (top) and $Q^2 = 100~\textrm{GeV}^2$ (bottom) for gluons, $\mathrm{\bar{u}}$, $\mathrm{u_V}$ and $\mathrm{d_V}$.}
\label{fig:protonPDF}
\end{center}
\end{figure}

The majority of the data in our proton baseline PDFs comes from the combined HERA collider DIS measurement \cite{H1:2015ubc} (1145 data points). In addition we also include some fixed-target DIS data from BCDMS \cite{BCDMS:1989qop} (327 data points) and NMC \cite{NewMuon:1996fwh} (100 data points) experiments. As an update with respect to our previous analysis we include also several EW-boson datasets measured in proton-proton collisions at the LHC (in total 132 data points). Here the NLO cross sections are computed using pre-calculated interpolation grids and NNLO corrections are obtained using $K$-factors. The relevant references are provided in Table 2 in \cite{Helenius:2021tof}. We find that when including LHC data in the baseline fit the resulting $\chi^2/N_{\mathrm{df}}$ value is somewhat smaller in our NNLO analysis (1.24) compared to the value in our NLO analysis (1.30). The resulting NNLO PDFs are compared to our earlier work in figure \ref{fig:protonPDF} at the initial scale of the analysis, $Q^2 = 1.69~\text{GeV}^2$, and at $Q^2 = 100~\text{GeV}^2$ for gluons, sea quarks, and $\mathrm{u}$- and $\mathrm{d}$-valence quarks. For all considered flavours the results are very similar to our earlier analysis but thanks to the added data the uncertainties are slightly reduced. We have also verified that the resulting proton PDFs provide a very good description of the recent EW-boson data from ATLAS \cite{ATLAS:2018pyl} that were not included in our analysis further validating the baseline.

\subsection{Nuclear PDFs}

\begin{figure}[htb]
\begin{center}
\includegraphics[width=0.32\textwidth]{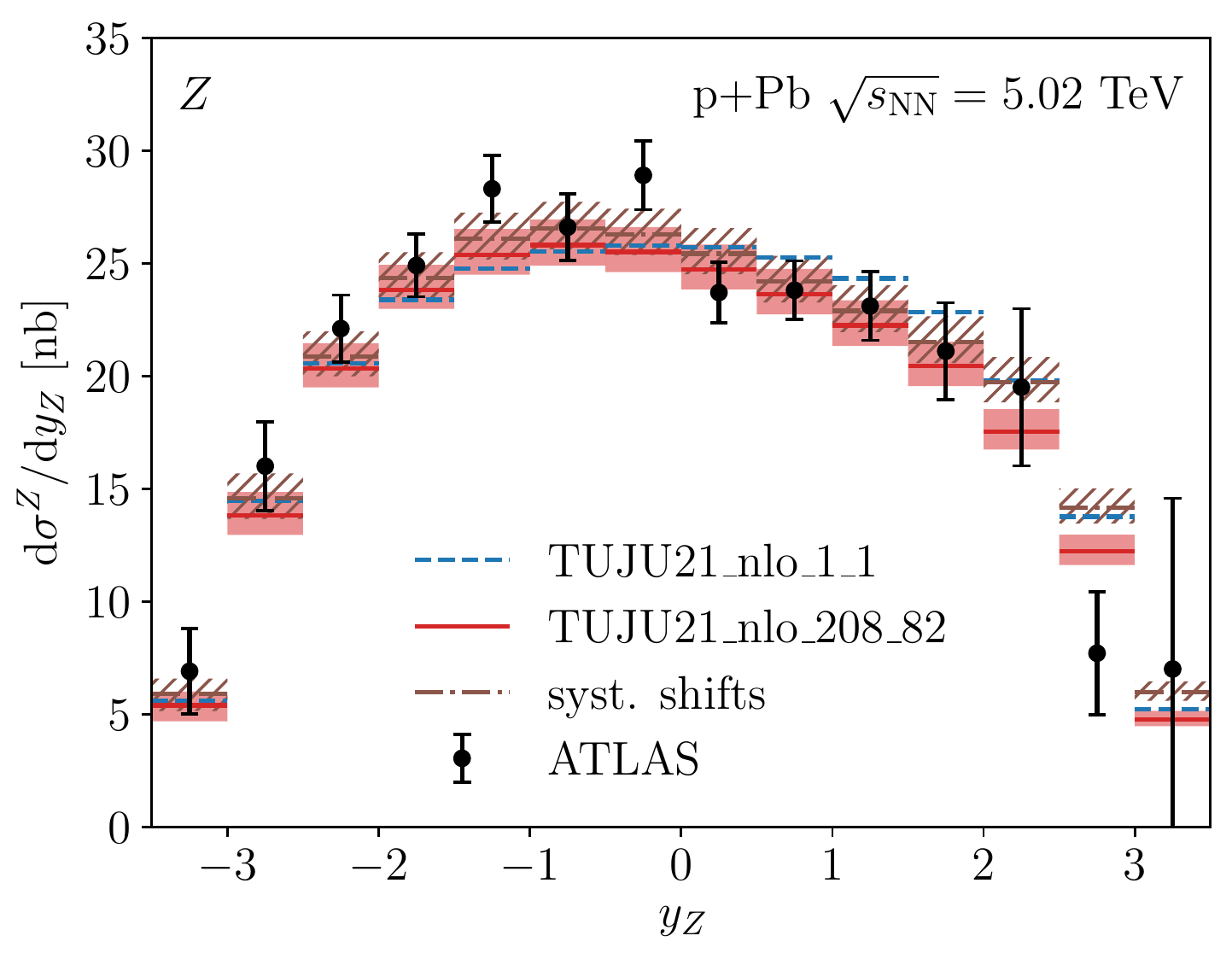}
\includegraphics[width=0.32\textwidth]{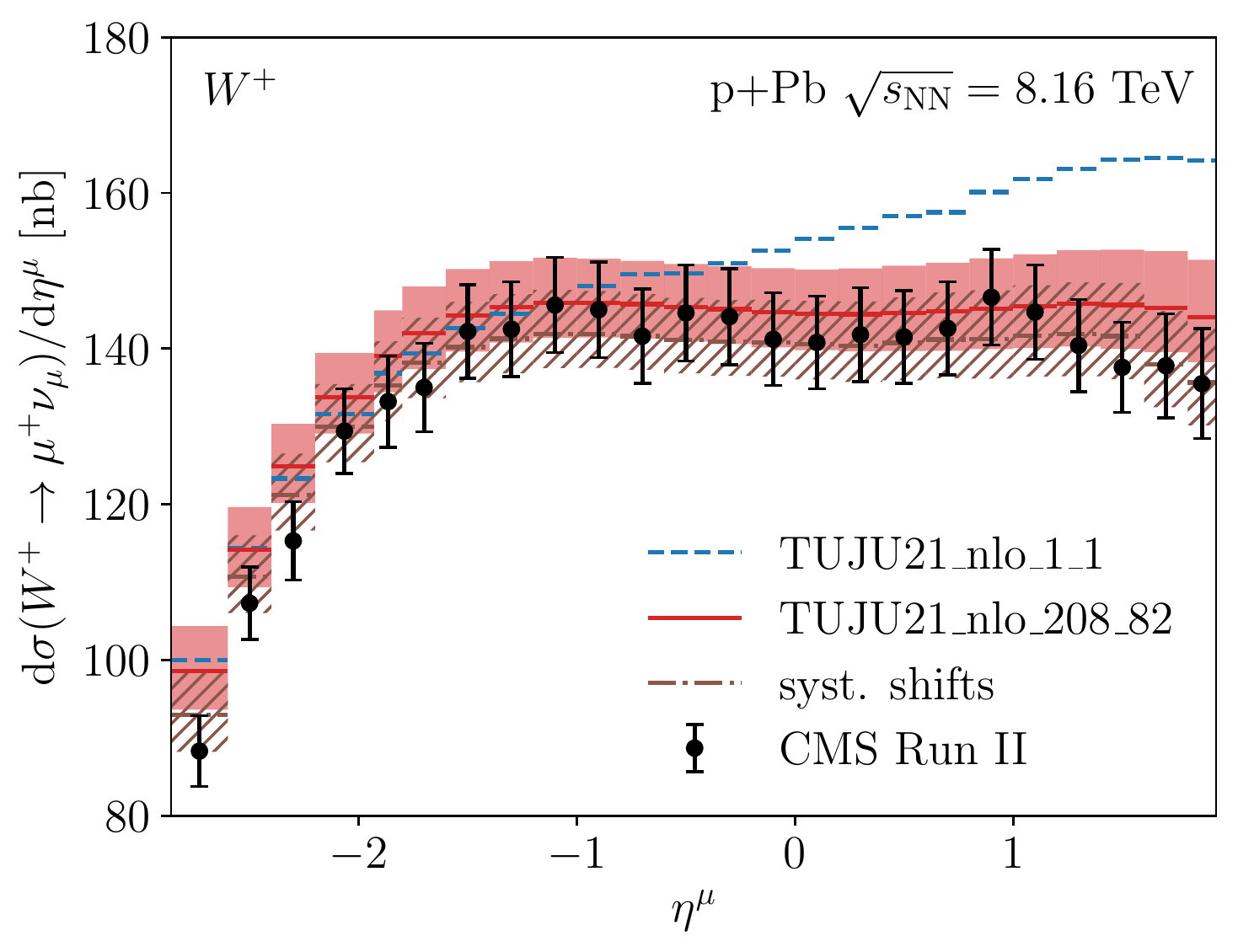}
\includegraphics[width=0.32\textwidth]{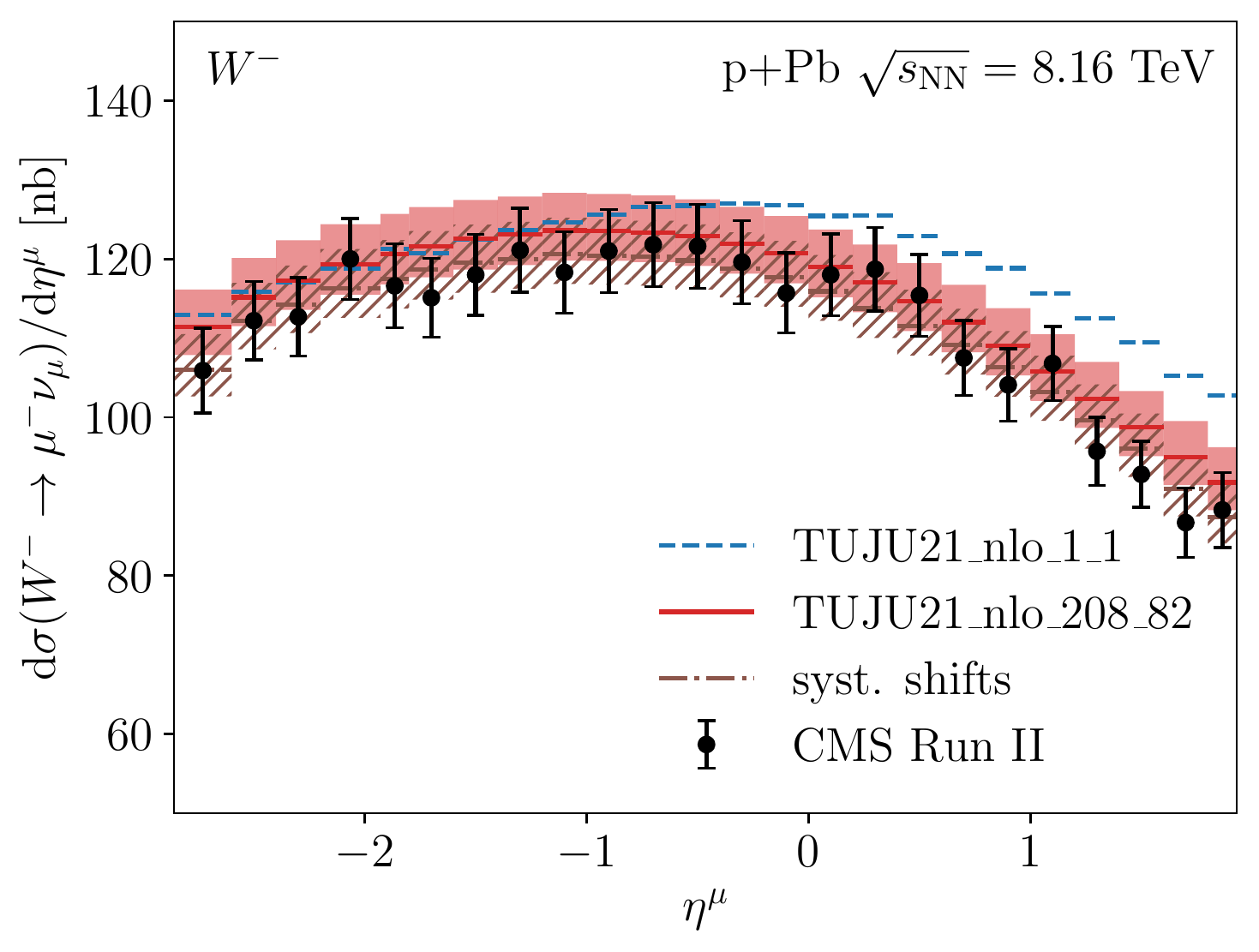}\\
\includegraphics[width=0.32\textwidth]{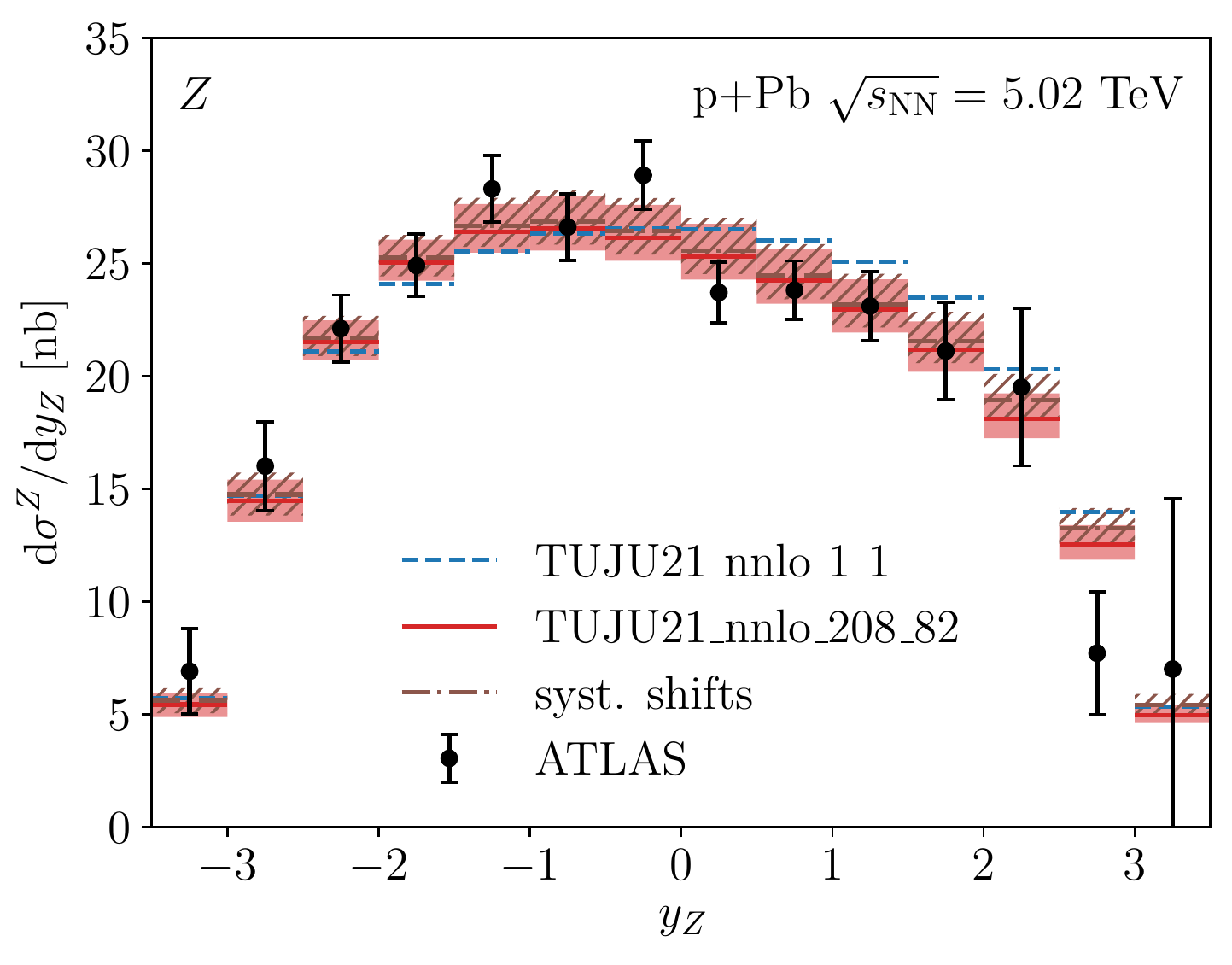}
\includegraphics[width=0.32\textwidth]{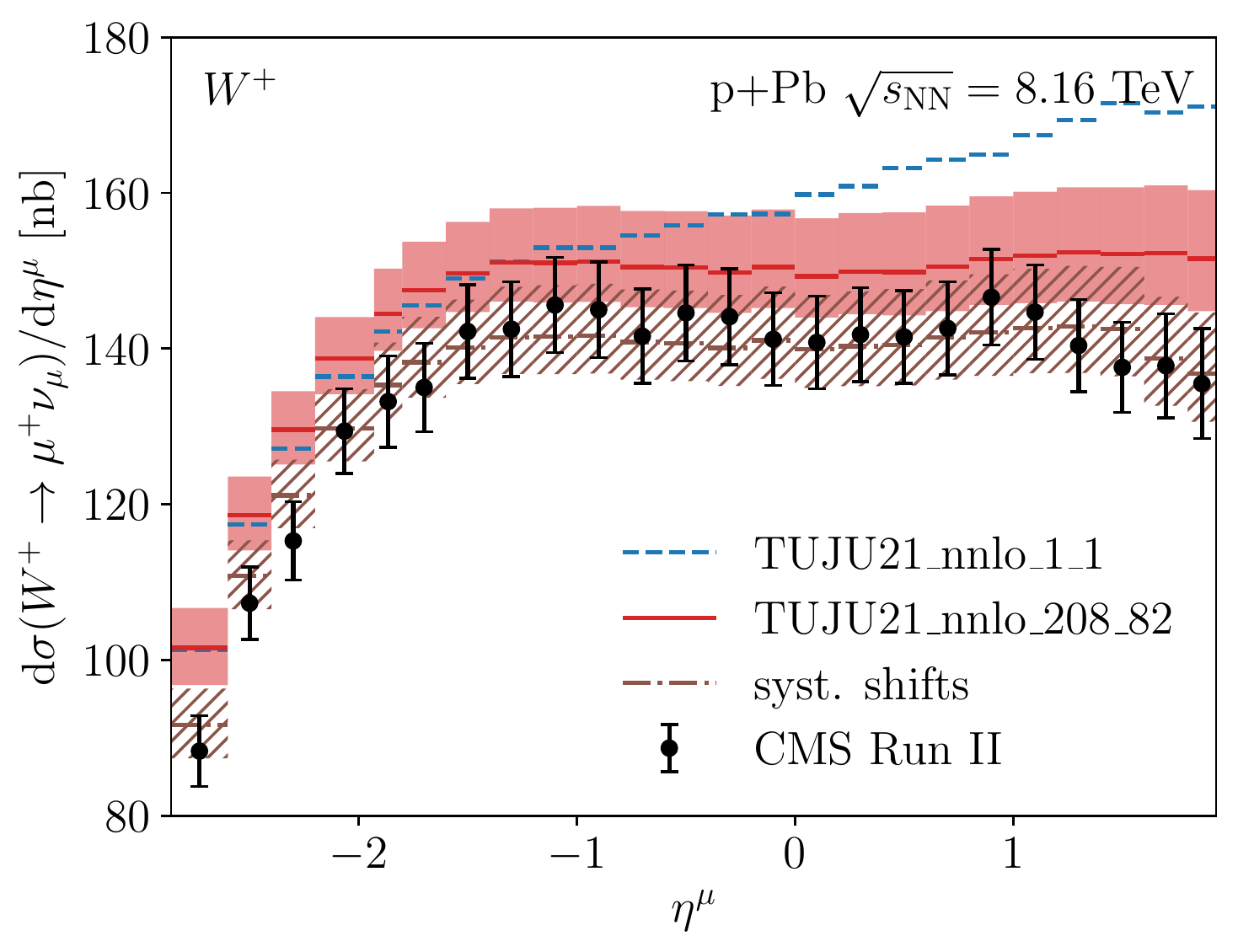}
\includegraphics[width=0.32\textwidth]{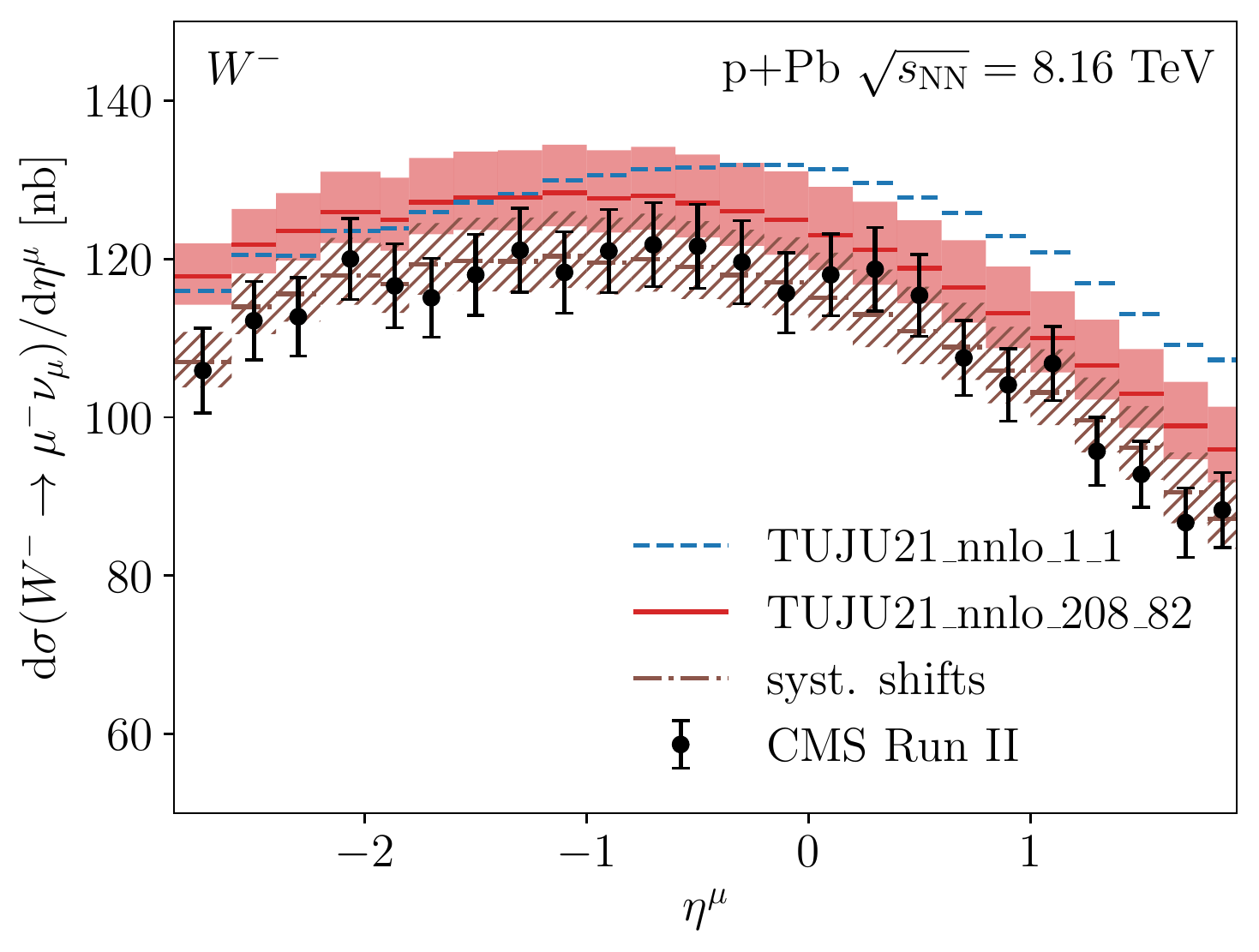}\\
\caption{Comparisons of the ATLAS data $Z$-boson and CMS $W^{\pm}$-boson production data in p+Pb collisions (black points) and the fit results. The computed cross sections include NLO (top) and NNLO (bottom) results without nuclear modification in the PDFs (dashed blue) and with (brown dot-dashed) and without (red solid) the fitted scaling factors.}
\label{fig:LHCdata}
\end{center}
\end{figure}

Our main results are the new nuclear PDFs applying EW-boson data from proton-lead collisions at the LHC (74 data points) on top of the fixed-target charged-lepton (600 data points) and neutrino DIS (1736 data points) data that were included earlier. With the nuclear data the NNLO interpolation grids were computed directly using MCFM version 8.3 \cite{Boughezal:2016wmq} without resorting to $K$-factors. Comparisons to ATLAS $Z$-boson \cite{ATLAS:2015mwq} and CMS $W^{\pm}$-boson \cite{CMS:2015ehw} data are presented in figure \ref{fig:LHCdata} for NLO and NNLO fits. Both fits provide a reasonable agreement with these data sets and especially the large-rapidity $W^{\pm}$ data favour clear nuclear suppression consistent with shadowing in nuclear PDFs. For the $W^{\pm}$ at NNLO the optimal agreement is obtained with somewhat larger shifts related to the correlated systematic uncertainties. In both cases the resulting shifts are, however, in line with the uncertainties quoted by the experiments. For the $Z$-boson production data the agreement with the nuclear PDFs is still better than without the nuclear modifications but the effect is not as clear as with the CMS $W^{\pm}$ data. Overall we find a very good agreement with the included data as indicated by the resulting $\chi^2/N_{\mathrm{df}}$ values, being 0.94 (NLO) and 0.84 (NNLO). The reduction of $\chi^2/N_{\mathrm{df}}$ from NLO to NNLO follows mainly from an improved description of neutrino DIS data and ATLAS $Z$-boson data.

\begin{figure}[htb]
\begin{center}
\includegraphics[width=\textwidth]{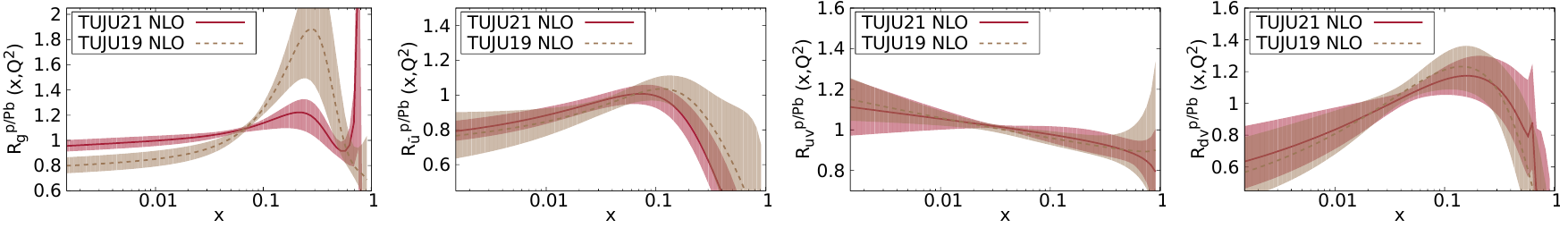}\\
\includegraphics[width=\textwidth]{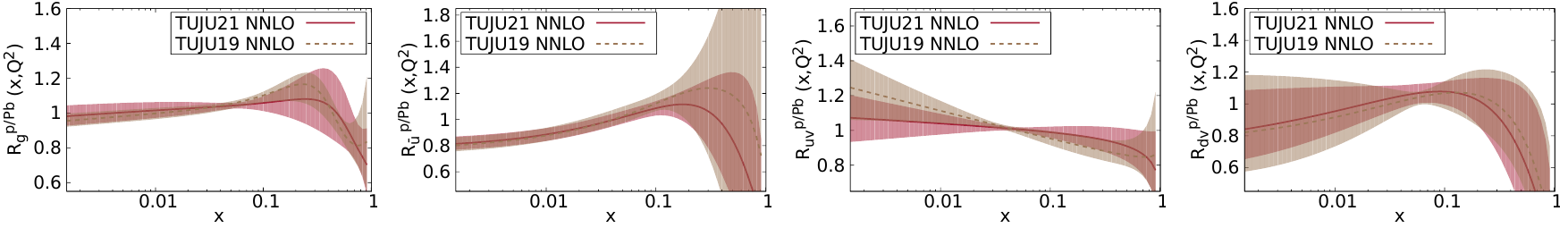}
\caption{Resulting nuclear modification ratios for our old TUJU19 and new TUJU21 NLO (top) and NNLO (bottom) analyses for gluons, $\mathrm{\bar{u}}$, $\mathrm{u_V}$ and $\mathrm{d_V}$ at scale $Q^2 = 100~\text{GeV}$.}
\label{fig:nuclearPDF}
\end{center}
\end{figure}

The resulting nuclear modification ratios for gluons, sea quarks, and $\mathrm{u}$- and $\mathrm{d}$-valence quarks in a proton bound to a lead nucleus are presented in figure \ref{fig:nuclearPDF}. The NLO and NNLO modifications are compared to our earlier TUJU19 analysis for each case at scale $Q^2= 100~\mathrm{GeV}$. The most prominent changes are the reduced gluon antishadowing around $x \approx 0.3$ and small-$x$ shadowing in the NLO fit and reduced sea quark uncertainties. The valence modifications of a bound proton are still mutually opposite, $\mathrm{d}$-valence preferring mild anti-shadowing enhancement around $x\approx 0.2$ whereas $\mathrm{u}$-valence showing slight suppression in the same region. However, when considering the average nucleon PDFs the opposite effects cancel to a large extent and the result is more in line with the other global analyses. Also, after adding the data the NLO and NNLO nuclear modifications are more in line than with only DIS data (TUJU19).

\section{Applications}

\begin{figure}[htb]
\begin{center}
\includegraphics[width=0.32\textwidth]{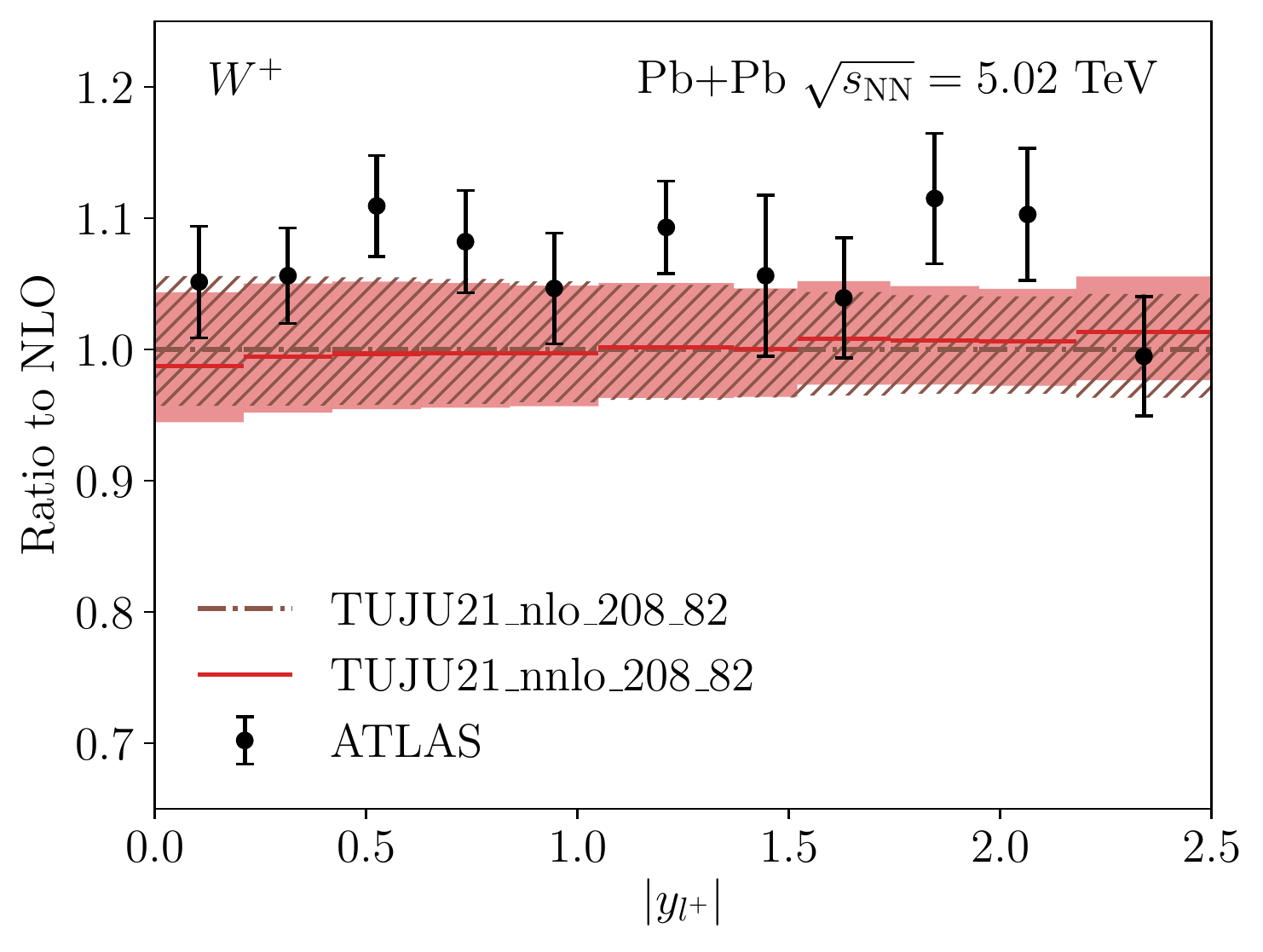}
\includegraphics[width=0.32\textwidth]{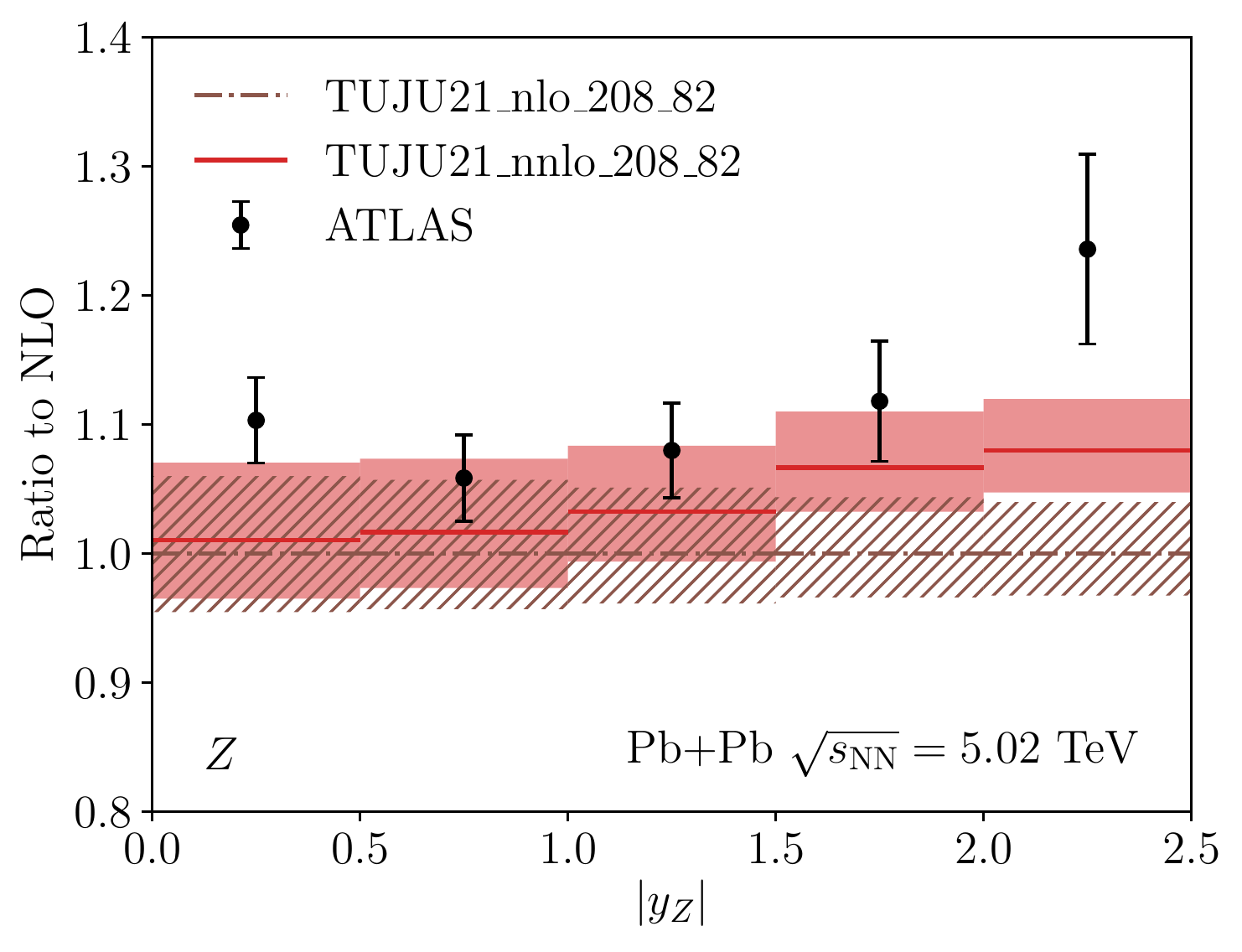}
\includegraphics[width=0.32\textwidth]{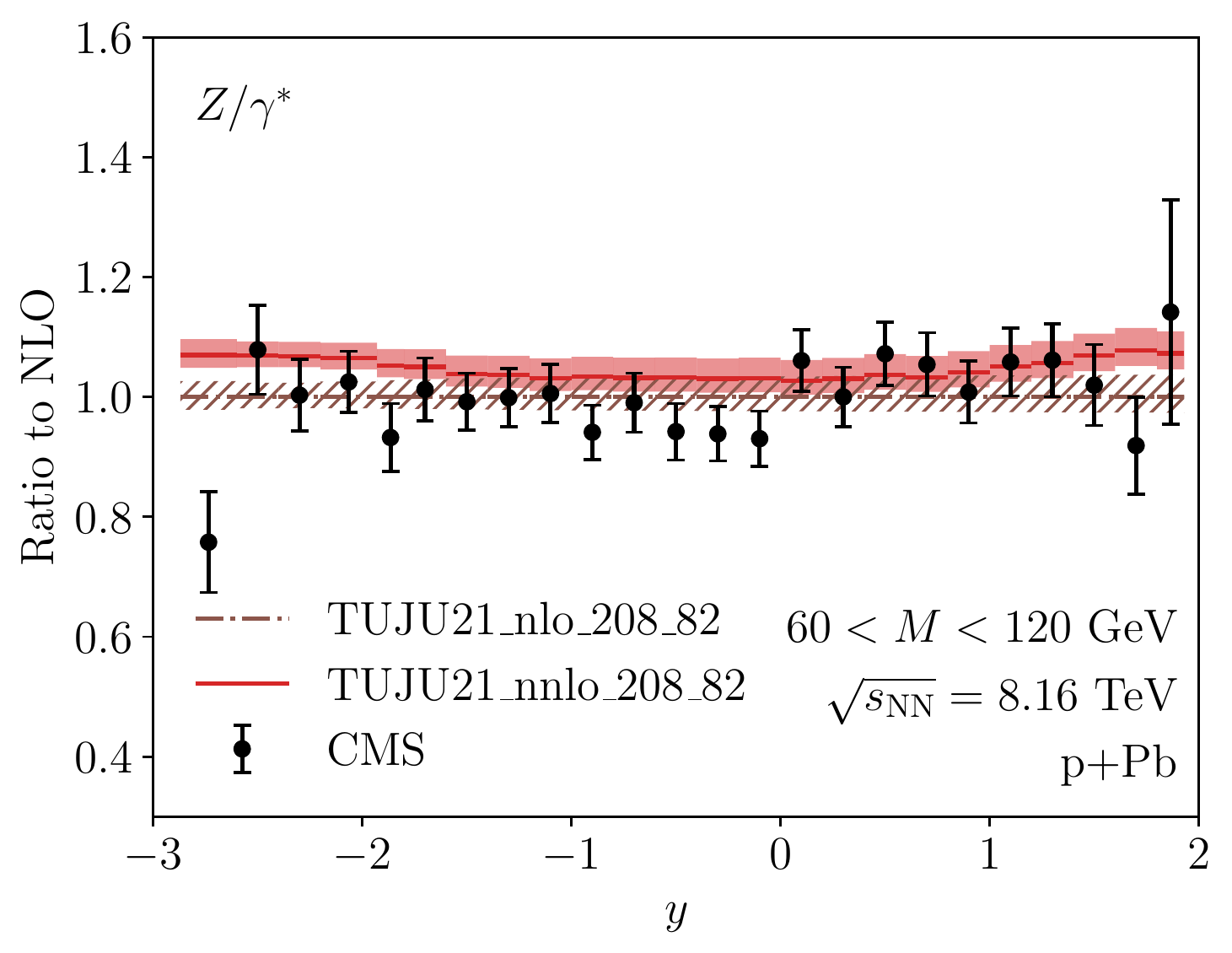}\\
\includegraphics[width=0.32\textwidth]{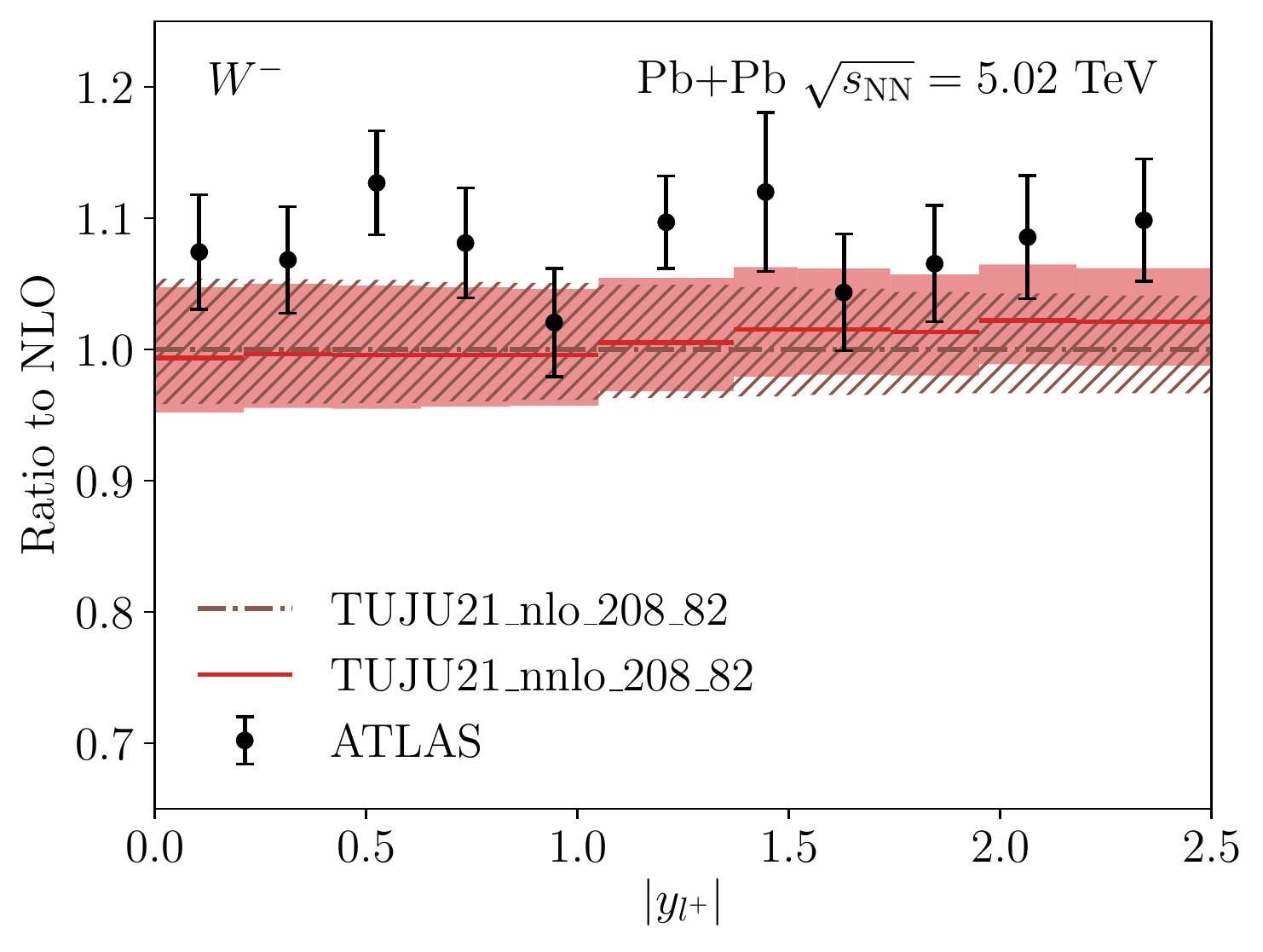}
\includegraphics[width=0.32\textwidth]{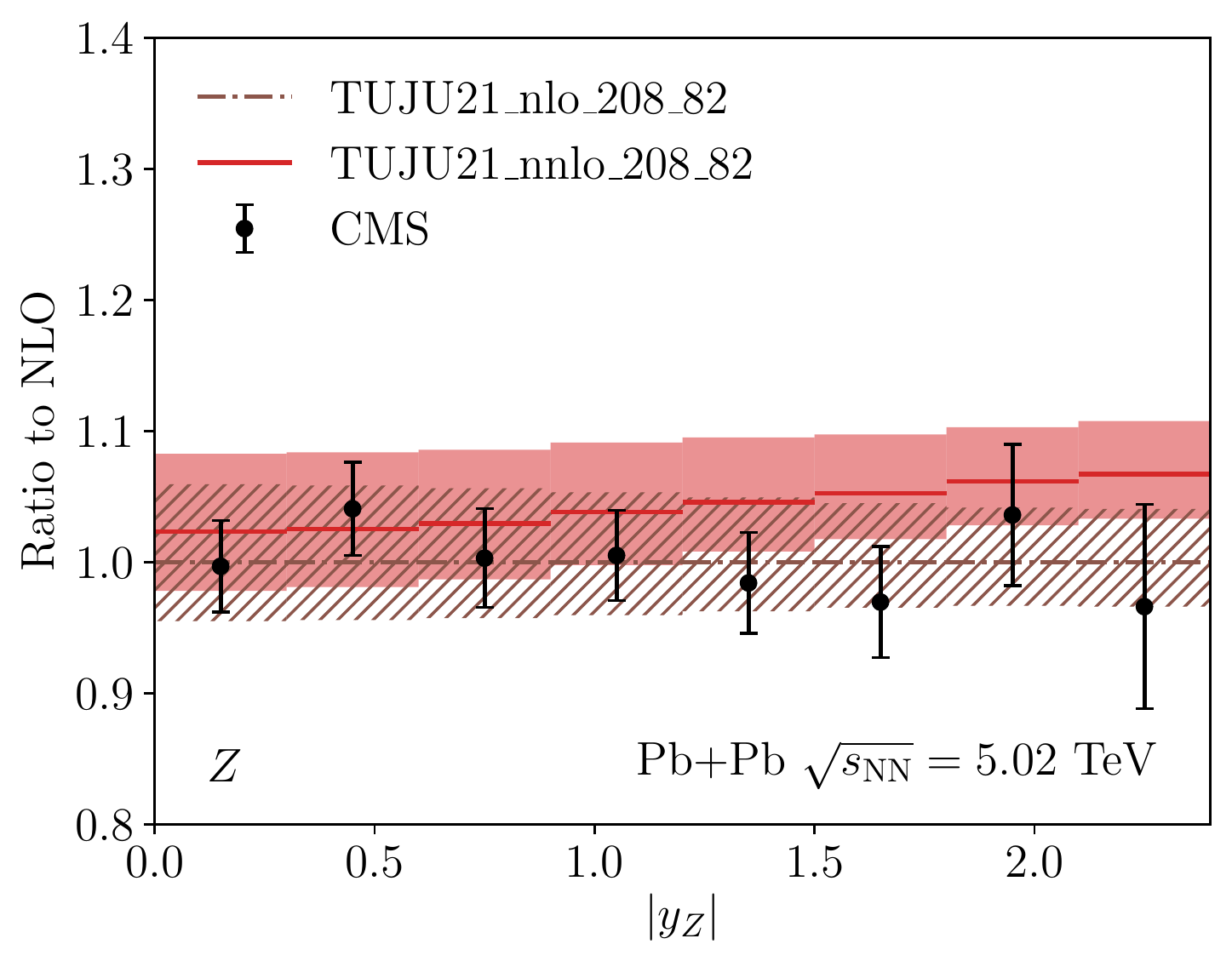}
\includegraphics[width=0.32\textwidth]{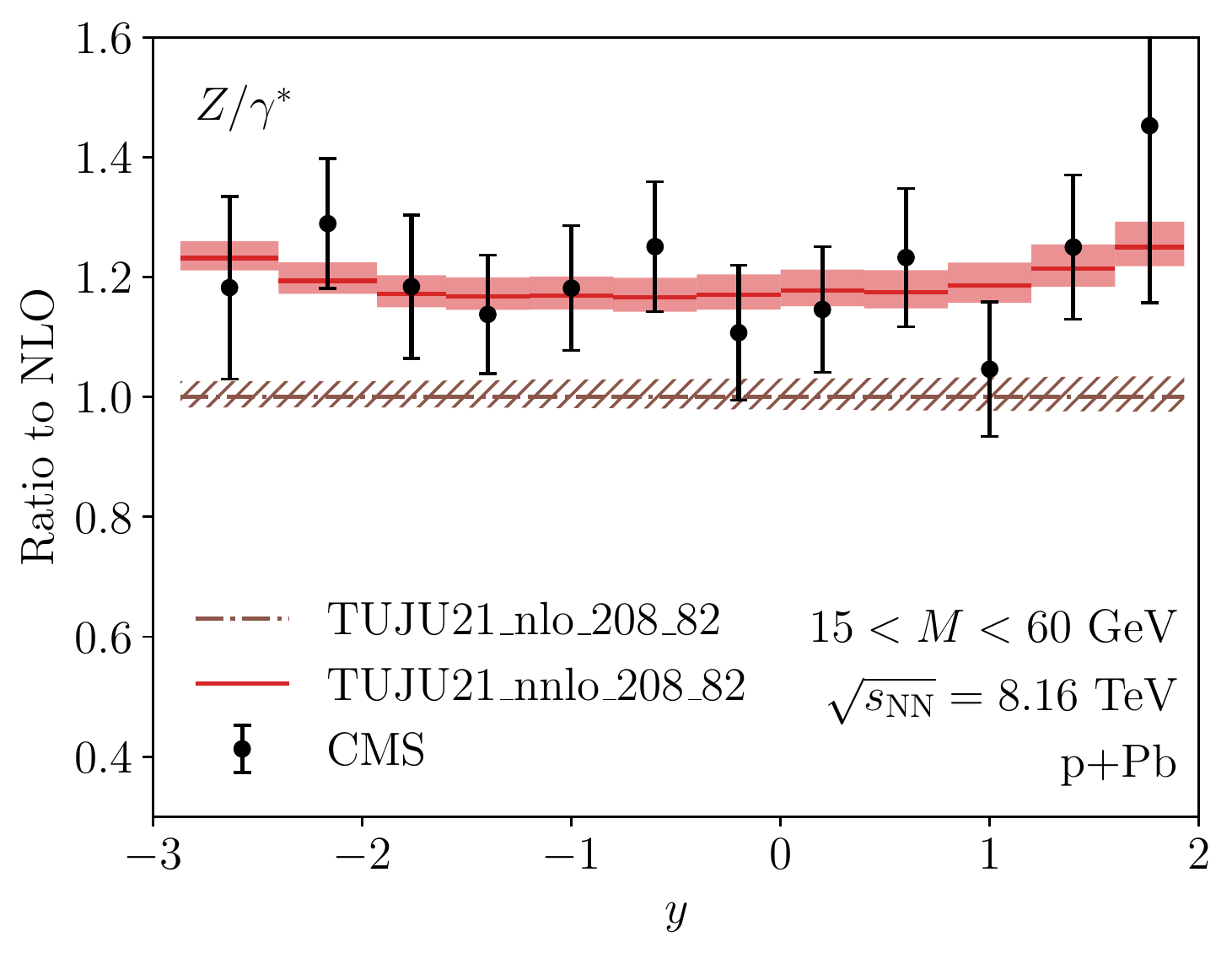}\\
\caption{Comparisons between the NLO and NNLO calculations and ATLAS data for $W^+$ (top left), $W^-$ (bottom left) and $Z$-boson production (top mid) in Pb+Pb collisions and to CMS data for $Z$-boson production (bottom mid). Also comparisons to the CMS DY measurement in p+Pb collisions for high-mass (top right) and low-mass bin (bottom right). All results are normalized to the NLO calculation.}
\label{fig:applications}
\end{center}
\end{figure}

As an application, we compare cross sections calculated with the resulting nPDFs at NLO and NNLO to the recent ATLAS and CMS measurements for EW-boson production in Pb+Pb collisions \cite{ATLAS:2019ibd, ATLAS:2019maq, CMS:2021kvd}, and to the recent Drell-Yan (DY) dilepton production data in p+Pb measured by the CMS \cite{CMS:2021ynu}. These comparisons are presented in figure \ref{fig:applications}. For all the ATLAS data the rapidity dependence is well reproduced in each case with the NNLO calculation but the normalization seems to be bit off. However, for the CMS data for $Z$-boson production it is actually the NLO calculation that describes the data better whereas the NNLO result is above the data hinting at a tension between the two datasets. The NNLO corrections for the considered observables in Pb+Pb collisions tend to be small. However, when considering the low-mass DY data the NNLO corrections become important and are larger than 20\% at the most extreme rapidities and it seems necessary to consider NNLO precision to match the data. 

\section{Summary}

We have presented a new nuclear PDF analysis, performed at NLO and NNLO accuracy in pQCD. It is the first NNLO analysis that includes also data from p+Pb collisions at the LHC. We find that the description of the data is significantly improved when going from NLO to NNLO and demonstrate that for some observables the NNLO corrections can be large. We conclude that with the increasing precision of the LHC data it will be necessary to make NNLO accuracy the default choice also for the nuclear PDFs.

\section*{Acknoledgements}

The authors acknowledge support by the state of Baden-Württemberg through bwHPC and also wish to acknowledge CSC-IT Center for Science, Finland, Project No. jyy2580. Also the support from the Academy of Finland, Projects No. 308301 and No. 331545, and through the Center of Excellence in Quark Matter, is acknowledged. This work was also supported in part by the Bundesministerium für Bildung und Forschung (BMBF), Grant No. 05P21VTCAA.

\end{document}